\documentclass[fleqn,usenatbib]{mnras}

\usepackage{newtxtext,newtxmath}

\usepackage[T1]{fontenc}

\DeclareRobustCommand{\VAN}[3]{#2}
\let\VANthebibliography\thebibliography
\def\thebibliography{\DeclareRobustCommand{\VAN}[3]{##3}\VANthebibliography}

\usepackage{graphicx}	
\usepackage{amsmath}	
\usepackage{hyperref}
\usepackage{subcaption}

\newcommand\HII{\ion{H}{II}~} 
\newcommand\HeI{\ion{He}{I}} 
\newcommand\HeII{\ion{He}{II}} 
\newcommand\HeIII{\ion{He}{III}} 

\usepackage{scalerel,tikz}
\usetikzlibrary{svg.path}
\definecolor{orcidlogocol}{HTML}{A6CE39}
\tikzset{orcidlogo/.pic={
 \fill[orcidlogocol] svg{M256,128c0,70.7-57.3,128-128,128C57.3,256,0,198.7,0,128C0,57.3,57.3,0,128,0C198.7,0,256,57.3,256,128z};
 \fill[white] svg{M86.3,186.2H70.9V79.1h15.4v48.4V186.2z}
 svg{M108.9,79.1h41.6c39.6,0,57,28.3,57,53.6c0,27.5-21.5,53.6-56.8,53.6h-41.8V79.1z M124.3,172.4h24.5c34.9,0,42.9-26.5,42.9-39.7c0-21.5-13.7-39.7-43.7-39.7h-23.7V172.4z}
 svg{M88.7,56.8c0,5.5-4.5,10.1-10.1,10.1c-5.6,0-10.1-4.6-10.1-10.1c0-5.6,4.5-10.1,10.1-10.1C84.2,46.7,88.7,51.3,88.7,56.8z};
}}
\newcommand\orcidicon[1]{\href{https://orcid.org/#1}{\mbox{\scalerel*{
\begin{tikzpicture}[yscale=-1,transform shape]
\pic{orcidlogo};
\end{tikzpicture}
}{|}}}}

\defcitealias{Wibking_2022}{WK23}
\defcitealias{Hu_2023}{HWK23}

\title[H$\alpha$ vs. YSOs]{On the limitations of H$\alpha$ luminosity as a star formation tracer in spatially resolved observations}

\author[Hu et al.]{
Zipeng Hu,$^{\orcidicon{0000-0002-3758-552X}\,1}$\thanks{zphu.charles@gmail.com (ZPH)}
Benjamin D.~Wibking,$^{\orcidicon{0000-0003-3175-2291}\,2}$
Mark R.~Krumholz$^{\orcidicon{0000-0003-3893-854X}\,1,3}$, and
Christoph Federrath$^{\orcidicon{0000-0002-0706-2306}\,1,3}$
\\
$^{1}$Research School of Astronomy and Astrophysics, Australian National University, Canberra, ACT 2611, Australia\\
$^{2}$Department of Physics and Astronomy, Michigan State University,East Lansing, MI,USA\\
$^{3}$ARC Centre of Excellence for Astronomy in Three Dimensions (ASTRO-3D), Canberra, ACT 2611, Australia\\
}

\date{Accepted XXX. Received YYY; in original form ZZZ}

\pubyear{2024}

\begin{document}
\label{firstpage}
\pagerange{\pageref{firstpage}--\pageref{lastpage}}
\maketitle

\begin{abstract}
This study examines the limitations of H$\alpha$ luminosity as a tracer of star formation rates (SFR) in spatially resolved observations. We carry out high-resolution simulations of a Milky Way-like galaxy including both supernova and photoionization feedback, and from these we generate synthetic H$\alpha$ emission maps that we compare to maps of the true distribution of young stellar objects (YSOs) on scales from whole-galaxy to individual molecular clouds ($\lesssim 100$ pc). Our results reveal significant spatial mismatches between H$\alpha$ and true YSO maps on sub-100 pc scales, primarily due to ionizing photon leakage, with a secondary contribution from young stars drifting away from their parent molecular clouds. On small scales these effects contribute significantly to the observed anti-correlation between gas and star formation, such that there is $\sim 50\%$ less anti-correlation if we replace an H$\alpha$-based star formation map with a YSO-based one; this in turn implies that previous studies have underestimated the time it takes for young stars to disperse their parent molecular clouds. However, these effects are limited in dense regions with hydrogen columns $N_\mathrm{H} > 3 \times 10^{21}$ cm$^{-2}$, where the H$\alpha$- and YSO-based SFR maps show better agreement. Based on this finding we propose a calibration model that can precisely measure the SFR of large molecular clouds (mean radius > 100 pc) with a combination of H$\alpha$ and CO observations, which provides a foundation for future study of star formation processes in extragalactic molecular clouds.
\end{abstract}




\section{Introduction}
\label{sec: introduction}
The H$\alpha$ emission line produced by hydrogen recombination is one of the most commonly used tracers for star formation, since it is emitted by the photoionized gas created by extreme ultraviolet photons from young massive stars. Observers routinely use this line to measure star formation rates (SFRs) for galaxies out to redshift $z \sim 2.5$, and on scales from whole-galaxy to kpc \citep[e.g.,][]{Kennicutt_1983, Shivaei_2015, Ellison_2017, Belfiore_2018}. There are two major advantages of this method: first, the H$\alpha$ line is easy to observe due to its brightness and location in the optical part of the spectrum, and because hydrogen is ubiquitous throughout the Universe; second, compared with other SFR tracers as the dust-reprocessed IR and UV continuum, H$\alpha$ traces only the most recent star formation ($< 5$ Myr) due to the short lifetimes of the stars that produce the bulk of the H$\alpha$ flux and the short recombination times of \HII regions \citep[e.g.,][]{Weisz_2012, Caplar_2019, Haydon_2020}.

Driven in part by the success of H$\alpha$-based galactic-scale SFR measurements, there has been great interest of extending such studies to the smaller scales characteristic of individual molecular clouds, $\lesssim 100$ pc. Measurements on such small scales are important for testing star formation theories under different conditions, and have been used to measure quantities such as the characteristic lifetimes of molecular clouds and the strength of stellar feedback \citep[e.g.,][]{Kruijssen14c, Kruijssen_2018, Chevance_2020, Chevance_2021}. Of all star formation tracers that are bright enough for routine use in extragalactic measurements, H$\alpha$ is the only one that traces star formation on time scales similar to cloud life times, which can be as short as $\sim 10$ Myr \citep[and references therein]{Chevance23a}. However, systematic uncertainties arise when converting H$\alpha$ luminosity into SFR on local regions around molecular clouds. The first uncertainty source is the dust attenuation effect, which can cause significant underestimation of local SFRs in regions of high extinction. However, as long as the extinction is not too severe, this effect can be calibrated out using the ratios of Balmer recombination lines (the Balmer decrement method) on both galactic and cloud scales \citep[e.g.,][]{Sanchez_2012}. The second problem is the stochasticity of massive star formation. In a $\lesssim 100$ pc region the SFR may be too small to sample all stellar masses and evolutionary states, which most often leads to underestimation of the SFR derived from the H$\alpha$ tracer \citep{da-Silva14b, Gregg24a}. The third and to date most poorly-calibrated bias depends on the size of \HII regions. As summarized in \cite{Kennicutt_2012}, such regions vary in scale from $< 100$ pc in the Milky way to $200-500$ pc in actively star-forming galaxies, implying that H$\alpha$-bright \HII region boundaries may be displaced away from the original star-forming sites. This effect is further complicated by non-uniform gas distributions around newborn stars, which allows ionizing photons to travel farther in directions with less gas and thus make the bias anisotropic \citep{Jin_2022}. 

There have been limited efforts to characterise and calibrate the effects of stochasticity and finite \HII region size. Observationally, one can compare H$\alpha$ to direct counts of newborn stars, or young stellar objects (YSOs), identified from IR observations. However, beyond the Milky Way such efforts are severely limited by resolution. To date the only published extragalactic comparison of these tracers is from the Large Magellanic Cloud \citep{Ochsendorf17a}, and analysis of these data together with comparable Galactic data sets indicates significant disagreements on cloud scales \citep{Krumholz_2019}. 

This situation leaves simulations as the best available tool for understanding the spatial scales on which the SFRs traced by H$\alpha$ begins to deviate from the true star formation rate as characterised by direct counts of YSOs, and for quantifying the size of the discrepancy. Although no theoretical effort has been made thus far to study this question, there have been several numerical works that try to calibrate other aspects of H$\alpha$ emission as a star formation tracer. \cite{Tacchella_2022} use simulations of Milky Way-like and dwarf galaxies to study the effect of dust attenuation and calibrate the relation between H$\alpha$ luminosity and SFR accounting for it, while \cite{Flores_Vel_zquez_2020} have estimated the time scales to which different SFR tracers are sensitive with cosmological simulations. However, such galactic scale models generally do not reach spatial resolutions of parsecs, or mass resolutions $\lesssim 100$ M$_\odot$, and thus can not resolve the star formation histories of individual molecular clouds. Kiloparsec-scale galactic box simulations with periodic boundaries can reach such resolutions \citep[e.g.,][]{Peters_2017, Kado_Fong_2020}, but lack a realistic large-scale galactic environment, which may affect molecular cloud density and spatial structure and ionizing photon transmission. For obvious reasons, they can not be used to study the reliability of H$\alpha$ over a full range of scales from galactic to sub-kpc. 

In this work, we perform high resolution MHD simulations to address these challenges. We start from a moderate resolution Milky-Way like spiral galaxy simulation to create a realistic galactic environment for molecular cloud evolution that we run to statistical steady-state, then increase the mass resolution and continue the simulation for a shorter period to resolve the structures on parsec scales. Our star formation algorithm stochastically draws stars from the initial mass function (IMF), traces the individual stellar evolution to calculate the feedback, and directly follows the propagation of ionizing photons through the galaxy using a Str\"omgren volume method. Via post-processing, we derive local chemical compositions, and generate synthetic CO, H$\alpha$, and YSO observations. Our goal is to answer the following questions: on what scales do star formation rates inferred from H$\alpha$ agree reasonably well with those inferred directly from YSOs? On scales where agreement breaks down, what are the physical mechanisms responsible? And can we calibrate the H$\alpha$ tracer to the YSO tracer to allow more accurate mesurements on the scales of individual molecular clouds?

The present paper is organized as follows: \autoref{sec: methods} summarizes the numerical method and simulation setup, then outlines the synthetic observation generation method. \autoref{sec: Results} compares our synthetic H$\alpha$ observation with observations of real galaxies in order to validate our methods, and presents scale-dependent comparisons of star formation rates inferred by H$\alpha$ versus by direct YSO counting. \autoref{sec: discussion} discusses the physical mechanisms behind the difference, develops a calibration method, and proposes possible future work in this area. \autoref{sec: conclusion} concludes the work done in this paper.

\section{Methods}
\label{sec: methods}
We describe in \autoref{sec: simulations} our Milky-Way like galaxy simulation, and list basic physical parameters of the simulated galaxy. We also describe the underlying physics and numerical methods. Then in \autoref{sec: synthetic observation} we introduce our method to produce synthetic SFR measurements traced by both H$\alpha$ luminosity and YSO counts.

\subsection{Simulations}
\label{sec: simulations}
Our galactic scale simulation is designed based on the whole-galaxy simulations by \citet[hereafter \citetalias{Wibking_2022}]{Wibking_2022} and \citet[hereafter \citetalias{Hu_2023}]{Hu_2023}. 
The simulation code we use is \textsc{GIZMO} \citep{Hopkins_2015, Hopkins_2016a, Hopkins_2016b}, which adopts a mesh-free, Lagrangian finite-mass Godunov method to solve the equations of magneto-hydrodynamics (MHD). The simulation include a time-dependent chemistry network that tracks the abundances of H~\textsc{i}, H~\textsc{ii}, \HeI, \HeII, and \HeIII, and uses these abundances to calculate hydrogen and helium cooling rates. It also includes metal line cooling computed by interpolating on a set of pre-computed CLOUDY \citep{Ferland_1998} tables assuming Solar metallicity and collisional ionization equilibrium, and as implemented by the GRACKLE chemistry and cooling library \citep{Smith_2017}. We implement algorithms for star formation, stellar feedback, and radiative cooling in the same way as in \citetalias{Hu_2023}. Here we provide a brief summary of the physics and numerical methods most relevant to this project, and refer the readers to \citetalias{Hu_2023} and \citetalias{Wibking_2022} for more details.

For every gas particle with a density $\rho_{\rm g}$ above a prescribed threshold $\rho_{\rm crit}$, we determine its local SFR density as
\begin{equation}
\rho_{\rm SFR} = \epsilon_{\rm ff}\frac{\rho_{\rm g}}{t_{\rm ff}},
    \label{eq: SFR}
\end{equation}
where $\epsilon_{\rm ff}$ is the star formation efficiency, and $t_{\rm ff}$ is the local gas free-fall time. Multiple observational methods across a wide density range yield a mean value of $\epsilon_\mathrm{ff}\approx 0.01$ \citep{Krumholz_2019}, with a dispersion about 0.2 dex in the Milky Way \citep{Hu_2022}. Therefore we adopt $\epsilon_{\rm ff} = 0.01$ in \autoref{eq: SFR}. The star formation density threshold we apply in our simulation is $\rho_{\rm crit} = 1000$ m$_{\rm H}$ cm$^{-3}$, chosen to produce approximate equality between the Jeans mass and the particle mass at our peak gas mass resolution $\Delta M = 89$ M$_\odot$ (see below). To avoid extremely small time steps in high density region, and the resulting high computational cost, we increase $\epsilon_{\rm ff}$ to 1 for gas particles with $\rho_{\rm g} > 100\rho_{\rm crit}$. In this way, we quickly convert high density gas particles into stellar particles, and set the highest density we resolve in this simulation to approximately $100 \rho_{\rm crit}$. During one time step $\Delta t$, gas particles with $\rho_\mathrm{SFR} > 0$ have a probability
\begin{equation}
P = 1- \text{exp} (-\epsilon_{\rm ff} \Delta t / t_{\rm ff})
    \label{eq: P}
\end{equation}
of being converted to a stellar particle. Once formed, the stellar particles represent sub-clusters that only interact with the galactic environment via gravity and stellar feedback.

When calculating the stellar feedback, we cannot integrate over the IMF because our stellar particles are too small: at our mass resolution, the expected number of SNe per particle is $\lesssim 1$, and the expected number of stars $\gtrsim 20$ M$_\odot$ -- the mass range that dominates ionizing photon production -- is $\ll 1$. Instead, for each stellar particle, we use the stellar population synthesis code \textsc{SLUG} \citep{Da_Silva_2012,Krumholz_2015} to stochastically draw its stellar population from a Chabrier IMF \citep{Chabrier_2005}. In this simulation we only include photonionization and type II supernova feedback, since they are the dominant mechanisms at the scale with which we are concerned. At each hydrodynamical time-step, \textsc{SLUG} reports for each star particle the instantaneous ionizing luminosity, $\dot{N}_{\rm ion}$, the number of individual type II supernovae, $N_{\rm SNe}$, and the mass of the ejecta, $M_{\rm ej}$, added during that time step; we derive the ionizing luminosities from a combination of Padova stellar tracks \citep{Schaller92a, Meynet94a, Girardi00a} and SLUG's ``starburst99'' treatment of stellar atmospheres \citep{Leitherer99a}, and draw our estimates of which stars end their lives in type II SNe, and the ejecta mass for those that do, from the tabulated results provided by \citet{Sukhbold_2016}. Assuming each supernova ejects an equal energy of 10$^{51}$ erg, we can easily derive the energy of the ejecta as $E_{\rm ej} = N_{\rm SNe} \times 10^{51}$ erg, and the momentum as $p_{\rm ej} = \sqrt{2E_{\rm ej}M_{\rm ej}}$. Then for type II supernovae feedback, we inject the ejected mass, momentum, and energy into the neighbouring gas particles proportional to the solid angle centered on the stellar particle and subtended by the effective intersection area between each nearby gas particle and the star, following the approach described in \cite{Hopkins_2018a, Hopkins_2018b}. As for the implementation of photonionization, we adopt the algorithm as in \cite{Hopkins_2018a}: for each star particle, we sort its nearby gas particles with increasing distance, and start from the first gas particle that has a temperature below 10$^4$ K, which we define as non-ionized. We then calculate the ionization rate $\Delta\dot{N}_{\rm ion}$ needed to fully ionize it, set the temperature of this gas particle to 10$^4$ K, and deduct $\Delta\dot{N}_{\rm ion}$ from $\dot{N}_{\rm ion}$. We repeat this process on the next closest non-ionized gas particle until $\dot{N}_{\rm ion} \approx 0$.


The initial conditions and setup of our simulation follow the same routine as described in \citetalias{Hu_2023}. Following that paper, we initially run the simulation for 0.7 Gyr at a mass resolution of $\Delta M = 859.3$ M$_\odot$. During this period the galaxy reaches a stable SFR close to $3 \; \rm M_{\odot}$ yr$^{-1}$. We then increase the mass resolution by a factor of 10. To do so, we replace each gas particle with 10 smaller particles randomly distributed around their parent particle's center of mass, with a distance of one fifth of the parental smoothing length. After the split, the child particle mass is divided by 10, the child smoothing length is divided by $10^{1/3}$, while all other physical parameters remain the same. In this way we improve the mass resolution to about 90 M$_\odot$; as noted above, for our cooling curve this is approximately equal to the Jeans mass at our star formation threshold, and the corresponding Jeans length is $\approx$ 1 pc, which we take to be the characteristic resolution of our simulation. This split snapshot becomes the initial condition for the next stage of our simulation, which we run for 23 Myr, which is about 4 times of the maximum age of massive stars that can emit ionizing photons ($\sim$ 5 Myr) and thus is long enough for the ionization pattern to reach a steady state at the new resolution. During this phase we output snapshots at a 1 Myr frequency. We show the face-on gas column density map $N_{\rm H}$ at 720 Myr (i.e., 20 Myr into the high resolution run) in the upper left panel of \autoref{fig: 4 maps}.

\subsection{Synthetic observation}
\label{sec: synthetic observation}

In order to study the difference between SFR measurements derived from H$\alpha$ and those using direct star counts, we need to produce two different synthetic maps from our simulation: a YSO mass distribution map and an H$\alpha$ luminosity map. In addition, we also produce a synthetic CO emission map, for comparison with real extragalactic molecular cloud observations.

\begin{figure*}
\begin{center}
\includegraphics[width=\textwidth]{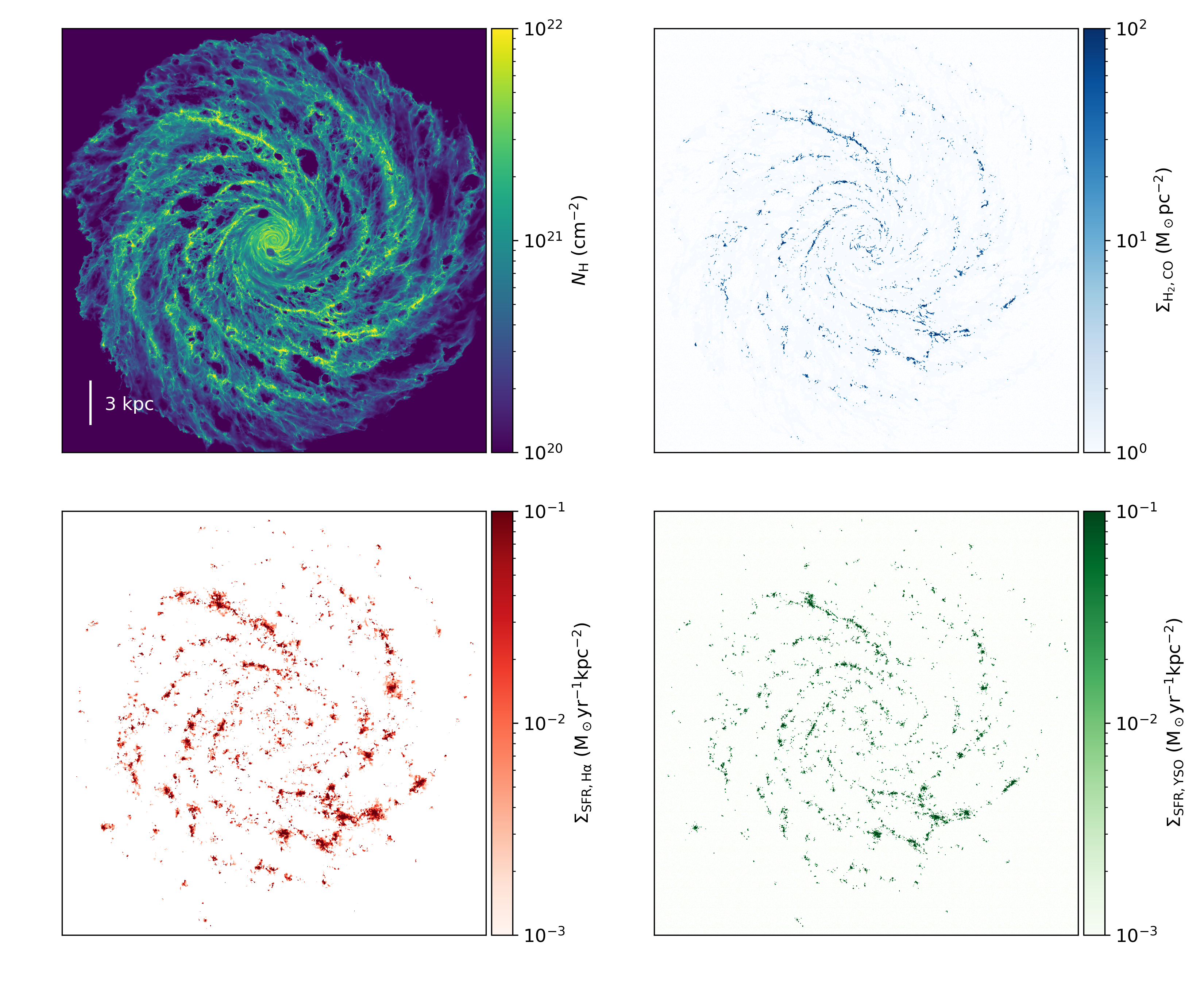}
\caption{Milky-Way-like spiral galaxy simulation at 720 Myr. Upper left panel: total gas column density map; here $N_\mathrm{H}$ is the column density of H nuclei, regardless of their chemical state. Upper right panel: surface density map of molecular mass as traced by CO $J = 1 \rightarrow 0$ emission (see main text for details). Lower left panel: SFR surface density map traced by H$\alpha$ emission. Lower right panel: SFR surface density map traced by YSOs with ages below 2 Myr. The dimensions of each map are $30\times30$ kpc, while the pixel size of each map is 20 pc. The ruler in the upper left panel indicates 3 kpc.}
\label{fig: 4 maps}
\end{center}
\end{figure*}

Due to their different evolution stages and spectral energy distributions, protostars identified from IR observations are normally classified into two types \citep[e.g.,][]{Pokhrel_2020}: class 1 YSOs are embedded in a circumstellar envelope, and have a smaller age limit ($\lesssim 0.5$ Myr), while class 2 YSOs are surrounded by an optically thick circumstellar disk but not an opaque envelope, and have a larger age limit ($< 2$ Myr). We elect to use a 2 Myr age limit to define YSOs for our analysis, both for consistency with existing extragalactic observational analyses \citep{Ochsendorf17a} and to reduce the shot noise that arises from the finite number of star particles in our simulation, which is limited by our mass resolution. To construct our YSO map, we identify all stellar particles formed in the 2 Myr prior to the time of a given snapshot as YSOs, and project their masses along the z-axis (normal to the galactic plane) onto a 2D histogram to produce the YSO mass distribution map; we defer for the moment a discussion of the pixel size we use in these maps. The SFR per unit area inside each pixel can be simply calculated as
\begin{equation}
\Sigma_{\rm SFR, YSO} = \frac{\sum m_{\rm YSO, pix}}{\Delta A_{\rm pix}\Delta t_{\rm YSO}},
\label{eq: SFR YSO}
\end{equation}
where $m_{\rm YSO, pix}$ is the total mass of YSOs inside a given pixel, $\Delta A_{\rm pix}$ is the pixel area, and $\Delta t_{\rm YSO} = 2$ Myr is the age range over which we identify YSOs. We plot the map of $\Sigma_{\rm SFR, YSO}$ in the lower right panel of \autoref{fig: 4 maps}, which illustrates the intrinsic spatial distribution of SFR in our simulated galaxy.

To construct the H$\alpha$ luminosity map, we first compute the case B H$\alpha$ emission coefficient from the approximation suggested by \citet{Draine_2011},
\begin{equation}
\alpha_{\rm eff, H\alpha} \approx 1.17 \times 10^{-13}T_4^{-0.942-0.031 \; \mathrm{ln}(T_4)}\mathrm{cm^3 s^{-1}},
\label{eq: H alpha coefficient}
\end{equation}
where $T_4$ is the ratio between the gas temperature $T$ and 10$^4$ K. Since we trace the abundances of hydrogen and helium atoms and ions in our simulation, we can easily derive the local number density of protons $n_\mathrm{p} = n_\mathrm{HII}$ and electrons $n_\mathrm{e} = n_\mathrm{HII} + n_\mathrm{HeII} + 2n_\mathrm{HeIII}$.\footnote{A minor technical issue is that, due to the difference between the time steps of \textsc{GIZMO} and \textsc{GRACKLE}, the chemical abundance of gas particles is not always updated to include the effects of photoionization when \textsc{GIZMO} writes out a snapshot. This results in a small number of particles that have been ionized by the photoionization algorithm and whose temperatures have therefore been set to $10^4$ K, but whose \HII abundances have not yet been updated. To handle these cases we treat all gas particles with $T \geq 10^4$ K as fully ionized in hydrogen, consistent with our photonionzation feedback algorithm.} Then we can derive the total H$\alpha$ luminosity from each gas particle simply as
\begin{equation}
L_{\rm H\alpha} = \alpha_{\rm eff, H\alpha} h \nu_{\alpha} n_\mathrm{e} n_\mathrm{p} V,
\label{eq: H alpha luminosity}
\end{equation}
where $\nu_{\alpha}$ = 457 THz is the frequency of the H$\alpha$ emission line, and $V$ is the effective volume of the gas particle. Since \autoref{eq: H alpha coefficient} is no longer a good approximation for low temperature regions, we discard $L_{\rm H\alpha}$ from gas particles with $T < 1000$ K. We also ignore $L_{\rm H\alpha}$ for gas particles with $\rho > 10 \rho_\mathrm{crit}$, since we are modifying the physics at these densities with our star formation algorithm and thus the particle properties are unreliable. To convert $L_{\rm H\alpha}$ to SFR, we consider a slightly modified version of the fit provided by \cite{Kennicutt_2012}, who find a relationship
\begin{equation}
    \mathrm{log \; SFR}(\mathrm{M_\odot\, yr^{-1}}) = \mathrm{log} \; L_{\rm H\alpha}(\mathrm{erg\, s^{-1}}) - C_\mathrm{H \alpha},
\end{equation}
where $C_\mathrm{H \alpha} = 41.27$ is the calibration parameter. This calibration includes the effects of dust scattering and attenuation, both external and internal to \HII regions (i.e., the fact that not every ionizing photon is absorbed by a hydrogen atom), which are not included in our simulation and post-analysis; it is also at least somewhat dependent on the choice of IMF and stellar evolution tracks to atmospheres. To ensure consistency, we therefore determine a different calibration factor as follows: for all snapshots past the first 3 Myr (during which time the high resolution simulation is still settling), we calculate both the total $L_{\rm H\alpha}$ and the total star formation rate traced by YSOs with ages $< 2$ Myr over the whole simulation domain, and set our value of $C_\mathrm{H\alpha}$ so that the star formation rates from H$\alpha$ and YSO counts agree when averaged over these snapshots. The resulting value is $C_\mathrm{H \alpha} = 40.67$, a factor of $\approx 3$ smaller than the \citet{Kennicutt_2012} value, which is consistent with the lack of dust attenuation in our simulations being the principle difference. We adopt this value throughout our analysis. We use this calibration to compute the H$\alpha$-derived star formation rate of every gas particle, and project the star formation rate converted from H$\alpha$ luminosity SFR$_\mathrm{H\alpha}$ along the z axis. We show a sample SFR surface density map from the 720 Myr snapshot in the lower left panel of \autoref{fig: 4 maps}.

The last mock observation we produce is the synthetic CO emission map, which is the most commonly used tracer of molecular mass. Our approach to this map is similar to that in \citetalias{Hu_2023}, so here we provide only a brief summary. While it is possible to include chemistry on the fly in simulations \citep[e.g.,][]{Glover_2010, Tritsis_2022}, this is computationally intensive, and also does not by itself capture non-LTE excitation effects, which are equally important for predicting observable emission. For this reason, we instead use the \textsc{DESPOTIC} astrochemistry and radiative transfer code \citep{Krumholz_2014_despotic} to post-process the output snapshots. Under the assumption of Solar elemental abundances and Solar radiation field environment, \textsc{DESPOTIC} uses an escape probability approximation to derive the CO $J = 1 \rightarrow 0$ emission line luminosity per hydrogen atom $l_\mathrm{CO}$ ($\nu_\mathrm{CO}$ = 115.271 GHz) as a function of three local physical parameters: hydrogen atom number density $n_\mathrm{H}$, column density $N_\mathrm{H}$, and velocity gradient $\mathrm{d}v/\mathrm{d}r$. We first run \textsc{DESPOTIC} on a grid covering the parameter range $n_{\rm H} = 10^{-2} - 10^{6} \; \rm cm^{-3}$, $N_{\rm H} = 10^{19} - 10^{25} \; \rm cm^{-2}$, and $dv/dr = 10^{-3} - 10^2 \; \rm km/s/pc$. The result is a table of $l_\mathrm{CO}$ versus $(n_{\rm H}, N_{\rm H}, \mathrm{d}v/\mathrm{d}r)$; we determine $l_\mathrm{CO}$ values for each gas particle by performing trilinear interpolation on this table. We take values of $n_{\rm H}$ and $\mathrm{d}v/\mathrm{d}r$ for each cell directly from \textsc{GIZMO} output, while we derive $N_{\rm H}$ from the $N_\mathrm{H}$ value at the gas particle position in the column density map. Then we can derive the total CO $J = 1 \rightarrow 0$ emission line luminosity as $L_\mathrm{CO} = X_\mathrm{H} m_\mathrm{g} l_\mathrm{CO}/m_\mathrm{H}$, where $X_\mathrm{H} = 0.76$ is the hydrogen mass fraction, and $m_\mathrm{g}$ is the gas particle mass. We set $L_\mathrm{CO} = 0$ for gas particles with ionized mass ratio $X_\mathrm{HII} > 10\%$; this is necessary because our DESPOTIC grid does not include the effects of photoionization. We refer the readers to \citetalias{Hu_2023} and \cite{Krumholz_2014_despotic} for more details. To facilitate comparison to observations, we convert $L_\mathrm{CO}$ into the CO-inferred molecular mass as $m_\mathrm{H_2, CO}/\mathrm{M}_\odot = 2.3\times 10^{-29}L_\mathrm{CO}/(\mathrm{erg\,s}^{-1})$; this conversion factor combines the CO $J = 1 \rightarrow 0$ line-luminosity conversion factor from \cite{Solomon_2005} and the mass-to-luminosity conversion factor from \cite{Bolatto_2013}. The resulting face-on synthetic $\Sigma_\mathrm{H_2, CO}$ surface density map for the 720 Myr snapshot is plotted on the upper right panel in \autoref{fig: 4 maps}.

\section{Results}
\label{sec: Results}
In this section, we first demonstrate that the spatial distribution of our synthetic H$\alpha$ luminosity map is similar to what is found in real observed galaxies by producing the so-called tuning fork diagram (\autoref{sec: tuning fork diagram}). Then we illustrate and quantify the differences between the SFR measurements traced by H$\alpha$ and YSOs as a function of scale (\autoref{sec: SFR tracers comparison}).

\subsection{Tuning fork diagram}
\label{sec: tuning fork diagram}
To test the reliability of our stellar feedback mechanisms and the resulting synthetic H$\alpha$ maps, especially on spatial scales of molecular clouds ($< 100$ pc), we need to demonstrate that our simulations are capable of reproducing the spatial decorrelation between H$\alpha$ emission peaks and molecular clouds that has been recently observed in a sample of nearby galaxies \citep{Kruijssen_2018, Kruijssen_2019, Chevance_2020, Chevance_2021, Kim_Jaeyeon_2022}. Reproducing this decorrelation has proven to be an exquisively sensitive test of feedback implementations in galaxy simulation codes \citep[e.g.,][]{Fujimoto19a, Semenov21a, Jeffreson21b, Jeffreson24a}.

Such decorrelation can be quantified by measuring the depletion time of molecular gas, defined as the ratio between the molecular mass traced by CO emission, and SFR traced by H$\alpha$ emission: $\tau_\mathrm{dep} = M_{\rm H_2,CO}/\rm SFR_{H\alpha}$. One measures $\tau_\mathrm{dep}$ in apertures of different sizes and centered on peaks of either the CO or H$\alpha$ map, then compare the results to the mean depletion time of the whole galaxy $\tau_\mathrm{dep, galaxy}$. Due to the spatial mismatch between CO and H$\alpha$ peaks, on sub-100 pc scales $\tau_\mathrm{dep} > \tau_\mathrm{dep, galaxy}$ when measured around CO peaks, and $\tau_\mathrm{dep} < \tau_\mathrm{dep, galaxy}$ when measured around H$\alpha$ peaks, but the difference diminishes as the aperture size increases, vanishing almost completely at $\sim$kpc scales.

\begin{figure}
    \includegraphics[width = \linewidth]{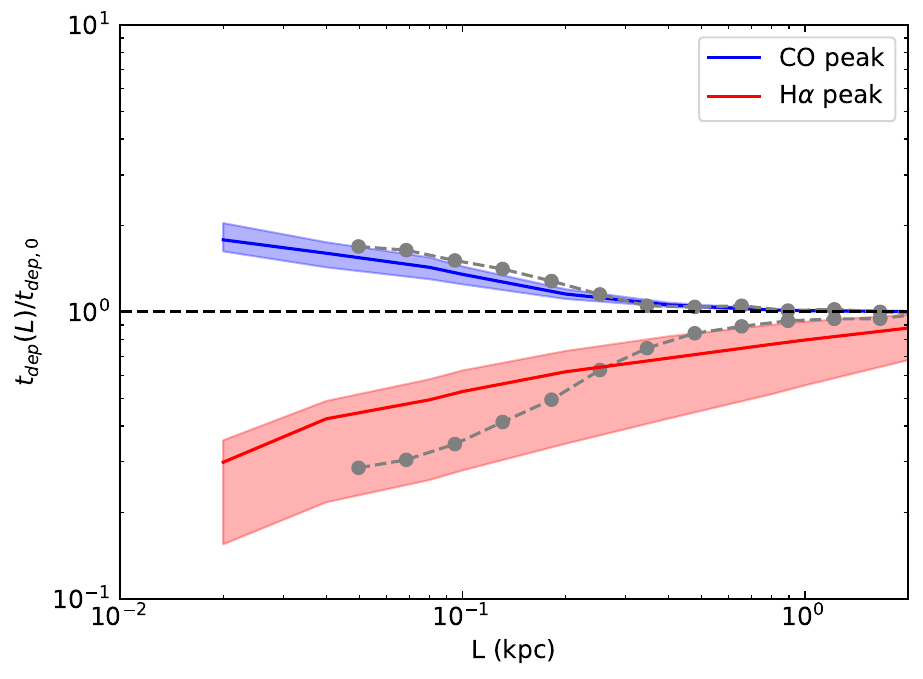}
    \caption{Tuning fork diagram from synthetic H$\alpha$ and CO emission observations. $\tau_\mathrm{dep, 0}$ is the mean molecular gas depletion time of the whole galaxy, while $\tau_\mathrm{dep}(L)$ is the depletion time calculated within an aperture of size $L$, centered around CO peaks (upper, blue branch) or H$\alpha$ peaks (lower, red branch). Both branches are stacked over 20 snapshots between 704 Myr to 723 Myr, with the solid line indicating the median and the shaded region showing the 16th to 84th percentiles. The dotted gray lines are the observed tuning fork diagram for NGC 628 \citep{Chevance_2020}.}
    \label{fig: tf plot}
\end{figure}

To mimic real observations, we first modify our maps by applying the resolution and detection limits reported by \citet{Kruijssen_2019} for the nearby spiral galaxy NGC 300, which are typical of other applications of this method. Their observations have a pixel size of 20 pc, a $\Sigma_\mathrm{H_2, CO}$ detection limit of 13 M$_\odot$ pc$^{-2}$, and a $\Sigma_{\rm SFR, H \alpha}$ detection limit corresponding to an H$\alpha$-inferred star formation rate $\Sigma_\mathrm{SFR,H\alpha} = 8.05 \times 10^{-3} \; \mathrm{M_\odot yr^{-1} kpc^{-2}}$. Then we follow the pipeline described by \cite{Kruijssen_2018} to generate the tuning fork diagram. The first step is to plot both $\Sigma_\mathrm{H_2, CO}$ and $\Sigma_\mathrm{SFR,H\alpha}$ maps at the finest available resolution of 20 pc (\autoref{fig: 4 maps}), then cut all pixels below the detection limits above. Applying these detection limit thresholds removes $\sim$ 11\% of the molecular mass and $\sim$ 13\% of the H$\alpha$ emission, so the effects on star formation and molecular mass are comparable. We then determine a mean depletion time $\tau_\mathrm{dep, 0}$ by summing the molecular gas mass and star formation rate over the remaining pixels and taking their ratio. Secondly, for a given scale $L$, we smooth both maps with a top hat filter with aperture size $L$, then identify local maxima from each map as emission peaks. If $L$ is large enough for some of the apertures to overlap, we randomly select 500 sub-samples of non-overlapping apertures around identified peaks, then calculate the mean $\Sigma_\mathrm{H_2, CO}$ and $\Sigma_\mathrm{SFR, H\alpha}$ values from the whole collection of sub-samples. The depletion time of aperture size $L$ ($\tau_\mathrm{dep}(L)$) is calculated as the ratio between these two mean values. Due to the stochasticity of our star formation algorithm, the birth of young massive stars is random in our simulation. To minimize this noise, we perform the analysis above for 20 snapshots between 704 Myr and 723 Myr, then plot the 16th, 50th and 84th percentiles of the ratio $\tau_\mathrm{dep}(L)/\tau_\mathrm{dep, 0}$ around CO emission peaks (top branch) and H$\alpha$ peaks (bottom branch) in \autoref{fig: tf plot}. We also show the measured values for NGC 628 \citep{Chevance_2020} for comparison; we choose this galaxy as an appropriate comparison because it has a total mass and size comparable to our simulated galaxy.

The plot shows that our simulated galaxy reproduces the observed tuning fork-like shape. At large scales ($L \leq $ 1 kpc), the two branches of the fork converge because the aperture size is large enough to include many CO peaks and H$\alpha$ peaks, making the depletion time equal to the galactic average value. When the aperture scale comes down, especially to $L < 100$ pc, the branches diverge as a result of decoupling between \HII region and molecular gas. The overall shape is in good agreement with the results from NGC 628 \citep{Chevance_2020}, and a quantitative comparison confirms this visual impression. Below in \autoref{sec: tuning fork effects} we apply the same \textsc{heisenberg} code \citep{Kruijssen_2018} that \citeauthor{Chevance_2020} apply to NGC 628 to estimate the characteristic spacing $\lambda$ between the CO emission peaks and H$\alpha$ emission peaks from our simulated images, and we obtain $\lambda = 103.2^{+58.3}_{-19.7}$ pc, whereas \citeauthor{Chevance_2020} find $\lambda = 113^{+22}_{-14}$ pc for NGC~628 -- identical to within the uncertainties. This confirms the reliability of our stellar feedback models, especially on cloud scale, and gives us the confidence to begin examining how star formation rate traced by H$\alpha$ differs from the ``true'' star formation rate one can infer directly from YSOs.

\subsection{Comparison of SFRs from H$\alpha$ emission and direct YSO counts}
\label{sec: SFR tracers comparison}
\begin{figure}
    \includegraphics[width = \linewidth]{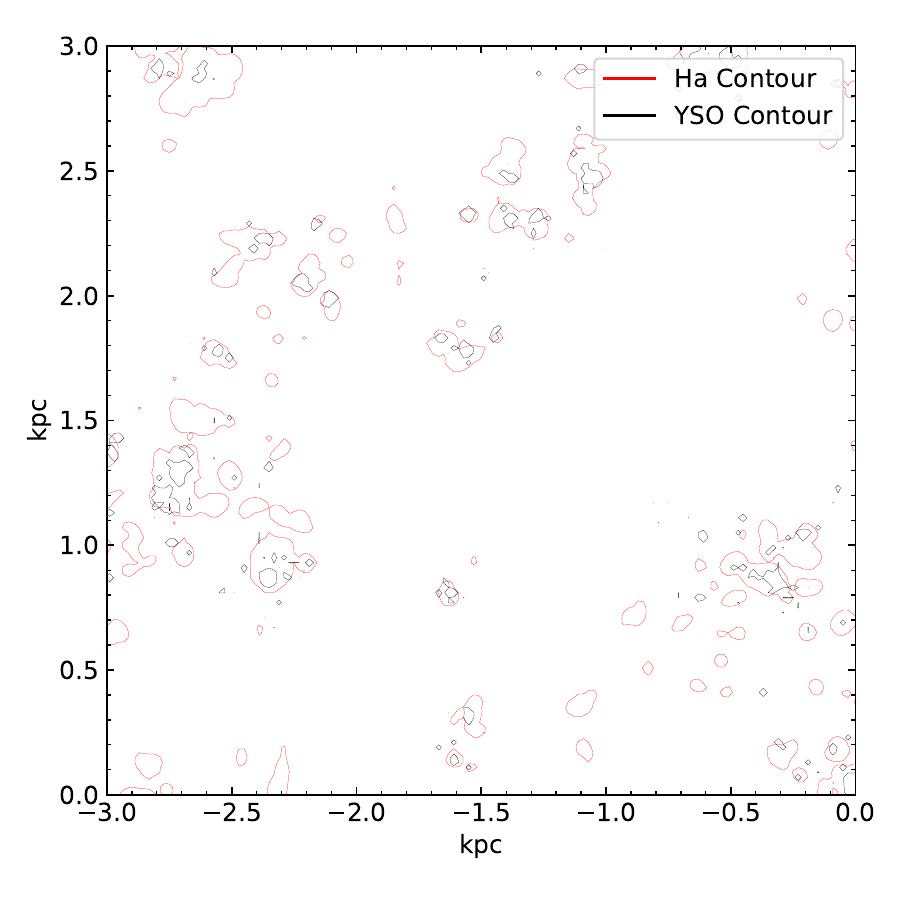}
    \caption{Contour plots of $\Sigma_{\rm SFR, H \alpha}$ (red) and $\Sigma_{\rm SFR, YSO}$ (black) within the inner galaxy. The contour level for $\Sigma_{\rm SFR, H \alpha}$ is $8.05 \times 10^{-3} \; \mathrm{M_\odot yr^{-1} kpc^{-2}}$, same as the detection limit in the observation of NGC 300; YSO contours are drawn out of pixels with at least one YSO. The galactic center is at the lower right corner of the plot, while the pixel size is 20 pc, the same as \autoref{fig: 4 maps}.}
    \label{fig: YSO Ha Contours}
\end{figure}

To study how well H$\alpha$ luminosity traces recent star formation history, we first compare their spatial distributions qualitatively. In \autoref{fig: YSO Ha Contours}, we plot contours of $\Sigma_{\rm SFR, H \alpha}$ (red) and $\Sigma_{\rm SFR, YSO}$ (black); the H$\alpha$ contour level corresponds to a typical detection limit of $8.05 \times 10^{-3} \; \mathrm{M_\odot yr^{-1} kpc^{-2}}$, and the $\Sigma_{\rm SFR, YSO}$ contour level is set to enclose pixels that include at least one YSO with age $<2$ Myr. To make the differences easier to see, we zoom in to a 3 kpc $\times$ 3 kpc patch of the galaxy, with the galactic center at the lower right corner; the pixel size is 20 pc for consistency. From the plot, we can clearly see a spatial mismatch: many \HII regions contain no YSOs (though the great majority of these do contain older stars, with ages of $2-5$ Myr), while others are much more spatially extended than the corresponding $\Sigma_{\rm SFR, YSO}$ contours. Averaged over 20 snapshots, the overlapping area between the two kinds of contours accounts for about only 38\% of the total area of H$\alpha$ contours, and about 70\% of the total area of YSO contours. Such mismatch indicates that on sub-100 pc scale, H$\alpha$ luminosity is an imperfect tracer of recent star formation history.

\begin{figure*}
\begin{center}
\includegraphics[width=\textwidth]{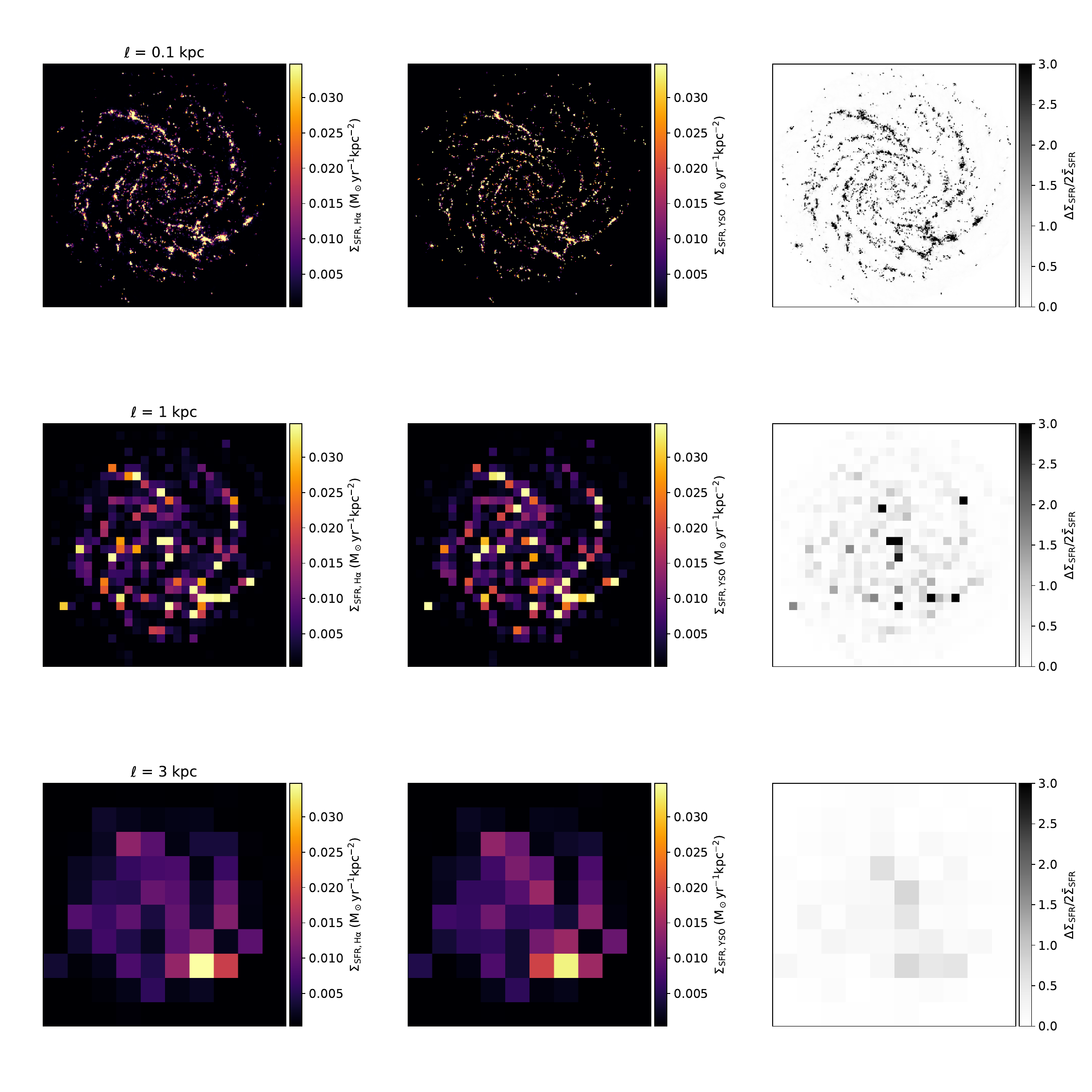}
\caption{SFR surface density $\Sigma_{\rm SFR}$ maps and their differences at different resolutions. All maps are generated from face-on projections of the simulation snapshot at 720 Myr. Plots in different rows uses different pixel sizes $\ell$, as indicated in the first column. The first two columns show the $\Sigma_{\rm SFR}$ maps traced by H$\alpha$ emission and YSOs, respectively, while the third column shows the normalised absolute difference between the first two columns (see main text for details).}
\label{fig: SFR diff maps}
\end{center}
\end{figure*}

To quantify the difference between the H$\alpha$- and YSO-based star formation rates as a function of spatial scale, we first construct maps of both $\Sigma_{\rm SFR, H \alpha}$ and $\Sigma_{\rm SFR, YSO}$ with pixel sizes $\ell$ ranging from 10 pc to 3 kpc. For each pair of maps with same pixel size, we define their relative difference as
 \begin{equation}
     \delta(\ell) = \frac{\sum_i \left|\Sigma_{{\rm SFR, H\alpha},i} - \Sigma_{{\rm SFR, YSO},i}\right|}{2 \sum_i \Sigma_{{\rm SFR, YSO},i}},
 \label{eq: SFR difference}
 \end{equation}
where $\Sigma_{{\rm SFR, YSO},i}$ and $\Sigma_{{\rm SFR, H\alpha},i}$ are the YSO- and H$\alpha$-derived SFRs in pixel $i$ for the map of pixel size $\ell$, and the sums run over all pixels. We illustrate this procedure in \autoref{fig: SFR diff maps}, where we plot the galactic $\Sigma_{\rm SFR}$ maps for H$\alpha$ and YSOs (1st and 2nd columns), and the normalised absolute difference $|\Sigma_\mathrm{SFR,H\alpha} - \Sigma_\mathrm{SFR,YSO}| / 2 \bar{\Sigma}_{\rm SFR,YSO}$ (3rd column), where $\bar{\Sigma}_{\rm SFR,YSO}$ is the mean of $\Sigma_\mathrm{SFR,YSO}$ over all pixels, at pixel sizes $\ell=0.1,$ 1, and 3 kpc (top to bottom rows). The quantity $\delta(\ell)$ defined in \autoref{eq: SFR difference} is simply the mean value of the maps in the third column. As shown in the figure, when $\ell$ increases, the difference between $\Sigma_{\rm SFR}$ maps drops, and thus $\delta$ becomes smaller.

\begin{figure}
    \includegraphics[width = \linewidth]{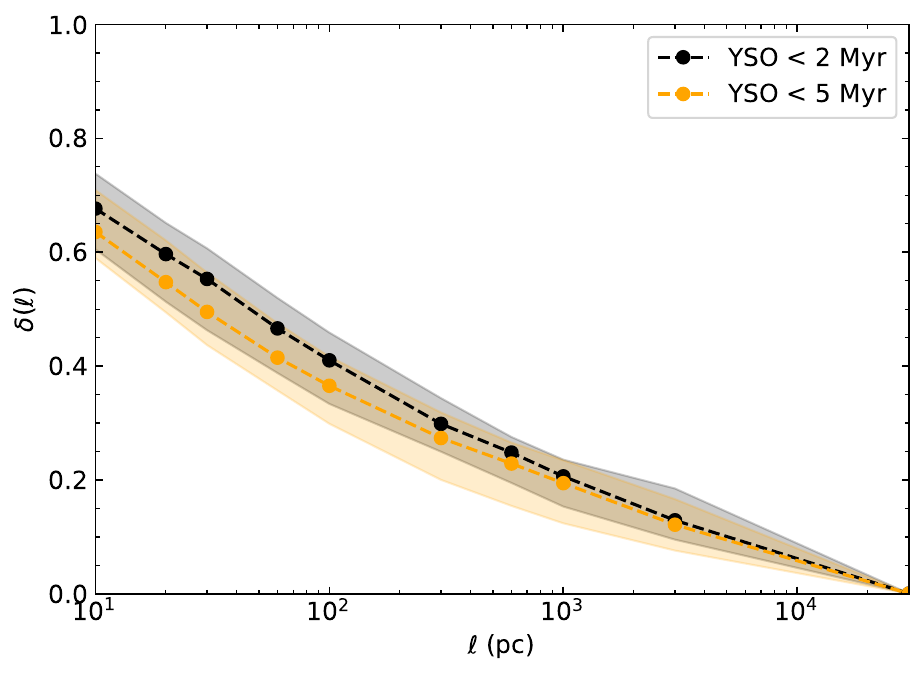}
    \caption{Normalised difference $\delta(\ell)$ between galactic SFR maps traced by H$\alpha$ and direct YSO counts as a function of pixel size $\ell$. The results are stacked from 20 snapshots from 704 Myr to 723 Myr. Different colors of the bands indicate different YSO age ranges used for SFR calculation: $< 2$ Myr (black) and $< 5$ Myr (orange). The dotted lines are the median values, while the band shows the 16th to 84th percentile range.}
    \label{fig: SFR diff}
\end{figure}

Because we have normalised the total SFRs derived from H$\alpha$ and YSOs to be equal when summed over the full galaxy, $\delta$ is constrained to lie in the range $0-1$, with $\delta = 0$ when the two maps are exactly the same and $\delta = 1$ when they are completely non-overlapping. We plot the $\delta(\ell)$ versus the pixel size $\ell$ in \autoref{fig: SFR diff}. Similar to \autoref{fig: tf plot}, we present the results stacked from 20 snapshots between 704 Myr and 723 Myr by showing the 50th percentile as the dotted lines, and the 16th and 84th percentiles as band edges. We also show the results from YSOs with over two ranges, for reasons we discuss below in \autoref{sec: why HII fail}. From $\ell = 1$ kpc to $\ell = 10$ pc, we see that $\delta(\ell)$ quickly increases, and reaching values $\sim 0.5$ at sub-100 pc scales. This indicates H$\alpha$ luminosity intrinsically fails to trace the recent star formation rate on sub-100 pc scales, and cannot be used for studying the star formation history of molecular clouds without a great deal of caution.

\section{Discussion}
\label{sec: discussion}
In this section, we first discuss the physical mechanisms behind the breakdown of H$\alpha$ as a SFR tracer on small scales (\autoref{sec: why HII fail}), and consider the implications of this for inferences of molecular cloud properties (\autoref{sec: tuning fork effects}). Then we develop an improved method to calibrate H$\alpha$ measurements for individual molecular clouds (\autoref{sec: calibration}). Finally we propose future work with the help of most recent observations (\autoref{sec: future work}).

\subsection{Why H$\alpha$ fails on small scales}
\label{sec: why HII fail}

We first seek to understand the physical mechanisms responsible for the breakdown of H$\alpha$ as a SFR tracer on small scales. As a start, we can immediately discard one obvious possibility, which is stochastic sampling of stellar populations \citep[e.g.,][]{Da_Silva_2012, da-Silva14b, Krumholz_2015, Arora21a}. While this is potentially an important effect at small scales, as we shall see below, stochasticity is not consistent with the pattern shown in \autoref{fig: SFR diff maps}. If the primary effect were stochasticity, we would expect to see many contours containing YSOs but little H$\alpha$, corresponding to regions containing a recently-formed population that is deficient in massive stars. While there are a few examples of such structures, \autoref{fig: SFR diff maps} shows that a far larger effect is H$\alpha$ contours that are more extended than YSO ones, or without any YSOs present at all.

This leads us to focus on two alternative explanations: drift of young massive stars, and transport of of ionizing photons. With regard to the former, we note that in Galactic observations YSOs identified from IR excess are normally younger than 2 Myr old, and we have followed this convention in our definition of YSOs. However, stars with significant ionizing luminosities can have lifetimes up to $\approx 5$ Myr. Once the stars are formed, they are no long subject to gas pressure and may drift out of their mother molecular cloud, and so one might hypothesise that H$\alpha$ differs from YSOs because it traces an older and more dispersed stellar population. We test for this effect in \autoref{fig: SFR diff}, where we compare the mismatch parameter $\delta(\ell)$ computed for our fiducial YSO age limit or 2 Myr (gray band) to one derived using an age limit of 5 Myr instead (orange band); the latter should include all significant sources of ionizing photons. We see that the extending age limit to 5 Myr does reduce the mismatch between H$\alpha$ and YSOs, but only by $\sim 10\%$. This suggests that stellar drift is not the primary culprit.

This leaves ionizing photon transport as the prime suspect. With regard to this phenomenon, we note that previous observations and theoretical models have found that ionizing photons can leak out of compact \HII regions, and create diffuse H$\alpha$ components up to even $\sim$ 1 kpc above the galactic plane \citep[e.g.,][]{Mathis_1986, Sembach_2000, Wood_2010, Belfiore_2022}. In the study of nearby disk galaxies by \cite{Pflamm_2008}, they find that this photon leak effect becomes significant for gas surface densities $\Sigma_\mathrm{gas} <10$ M$_\odot$ pc$^{-2}$. However, their $\Sigma_\mathrm{gas}$ measurements are averaged over $\sim$kpc scales, so the photon leak effect on cloud scales is still not well quantified. Our simulation, however, is perfectly suited to such a study, because when a gas parcel is newly ionized we set its temperature to exactly $10^4$ K, and therefore gas parcels at this temperature perfectly outline the outer boundaries of \HII regions. We can easily locate these particles in the output snapshots, allowing us to determine how far ionizing photons can travel.

\begin{figure}
    \includegraphics[width = \linewidth]{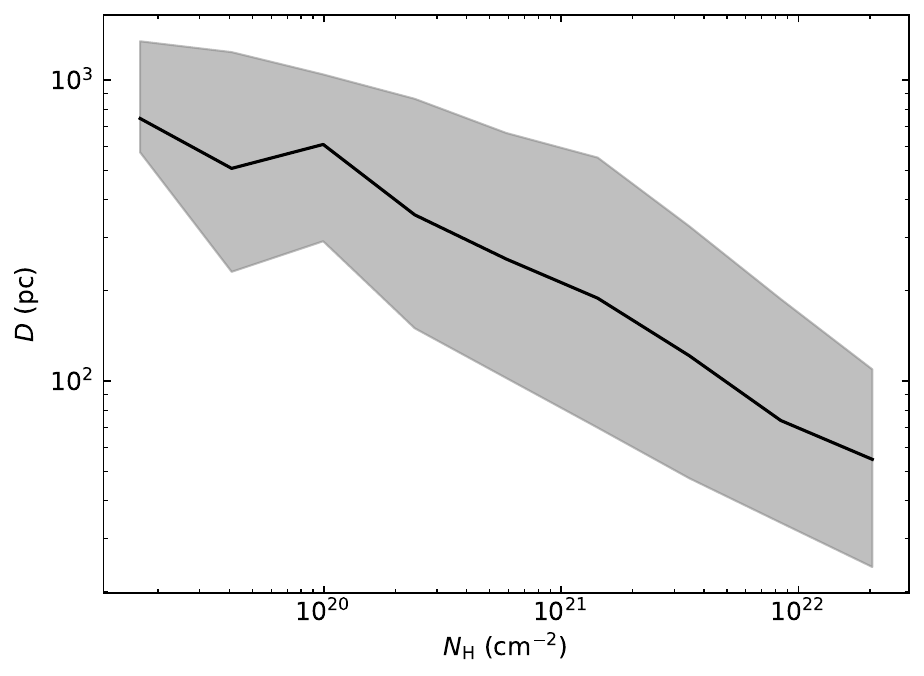}
    \caption{\HII region size $D$ versus the column density around stars $N_\mathrm{H}$. The solid line shows the median value, while the top and bottom edges of the shaded region are the 16th and 84th percentiles. $D$ is measured as the distance between gas particles that are just photonionized and the closet star born within 5 Myr, while $N_\mathrm{H}$ is measured inside a beam with a diameter of 100 pc around that star. The results are stacked from galactic maps of 20 snapshots.}
    \label{fig: photon Dis}
\end{figure}

Our analysis comes in three steps: first we record the coordinates of all gas particles with temperature of exactly $T = 10^4$ K; then we build a \textsc{k-d} tree containing the coordinates of stars with ages $<5$ Myr, which are potential sources of ionizing photons; finally, for each identified gas particle, we find the closet young star using the \textsc{k-d} tree. For each gas particle, we record both the distance $D$ to the closest star and the total gas column density $N_\mathrm{H}$ around the star; we measure $N_\mathrm{H}$ in a circular region around the star with a radius of 100 pc in order to smear fluctuations while retaining cloud scale information. We perform this analysis on snapshots from 704 Myr to 723 Myr, and we plot the 16th, 50th, and 84th percentiles of $D$ versus $N_\mathrm{H}$ in \autoref{fig: photon Dis}. This plot shows a clear trend that $D$ decreases with $N_\mathrm{H}$, which indicates that denser regions can absorb more ionizing photons, and confine the \HII regions to smaller sizes. The median \HII region size drops below 100 pc for $N_\mathrm{H} > 2 \times 10^{21}$ cm$^{-2}$, suggesting that we should find better agreement between H$\alpha$ and YSO-based SFRs if the YSOs are embedded in dense regions. 

\begin{figure}
    \includegraphics[width = \linewidth]{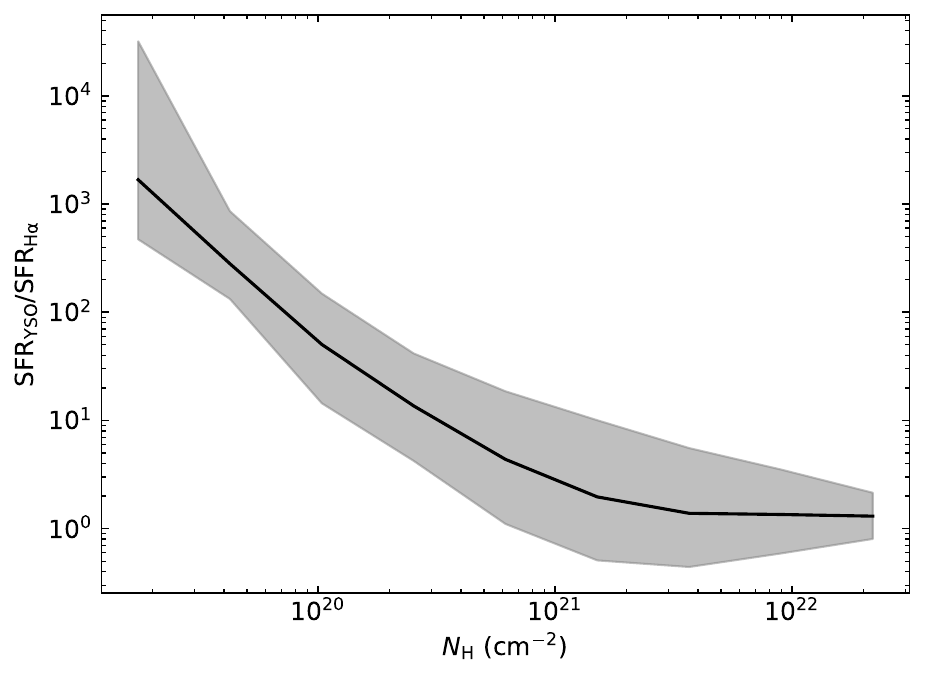}
    \caption{Ratios between SFRs traced by H$\alpha$ luminosity and YSO versus the column density $N_\mathrm{H}$, both calculated from 100-pc-large pixels with YSOs inside. The results are stacked from 20 snapshots between 704 Myr and 723 Myr. The mid solid line is the median value, while the shaded region shows the range between the 16th and 84th percentiles.}
    \label{fig: SFR ratio vs NH}
\end{figure}

To test this hypothesis, we plot the galactic SFR maps from the same snapshots with a pixel size of 100 pc; then for each pixel containing at least one YSO, we calculate the ratio $\Sigma_{\rm SFR, YSO}/\Sigma_{\rm SFR, H\alpha}$ and the column density $N_\mathrm{H}$. We plot this ratio as a function of $N_\mathrm{H}$ in \autoref{fig: SFR ratio vs NH}, where we again show the median value as the solid line and the 16th to 84th percentiles as the shaded region. As the Figure shows, when YSOs reside in a low surface density region, the corresponding \HII region greatly expands beyond the 100 pc pixel, leading to a substantial mismatch between the H$\alpha$ and YSO-based SFRs. Conversely, when the local column density becomes high enough ($N_\mathrm{H} \gtrsim 3 \times 10^{21}$ cm$^{-2}$, or $\Sigma_\mathrm{gas} > 32 \; \mathrm{M_\odot \; pc^{-2}}$), H$\alpha$ begins to agree very well with YSOs because the \HII region is well confined. We therefore conclude that H$\alpha$ luminosity is only locally trustworthy for dense regions with $N_\mathrm{H} \gtrsim 10^{21}$ cm$^{-2}$, while the exact threshold needs calibration from future high-resolution observations.

Finally, we note that while we have shown above that drift of young stars away from their birth sites is not the direct cause of the H$\alpha$-YSO mismatch, it may play a significant role in enhancing ionizing photon transport, together with the closely related effect that young stars drive away nearby dense gas with pre-supernovae stellar feedback. In either case, the effect is that the local column density around the stars drops with time, which enables the ionizing photons to travel further and enlarges the SFR tracers mismatch. Quantitatively, we find that the column density around a star on average drops by a factor of $\approx 10$ from the moment of its birth to the age of 5 Myr. Moreover, about 71\% of the YSOs $< 2$ Myr old reside in CO contours where $\Sigma_\mathrm{H_2, CO} \geq 13$ M$_\odot$ pc$^{-2}$, our rough threshold for \HII region confinement to $\lesssim 100$ pc sizes; for stars with ages between 2 and 5 Myr, this fraction drops to $\approx 38\%$. Thus the mismatch between H$\alpha$ and YSOs due to ionizing photon transport is mainly driven by older YSOs that have, either by drift or by dispersing their natal gas, escapes to low-density regions from which their ionizing photons can propagate long distances. This phenomenon has also been observed in another recent theoretical modeling effort by \citet{McClymont_2024}, who find that the diffuse ionized gas (DIG) is primarily ionized by older stars instead of embedded YSOs. 


\subsection{Implications for statistical inferences of the molecular cloud life cycle}
\label{sec: tuning fork effects}

\begin{figure}
    \includegraphics[width = \linewidth]{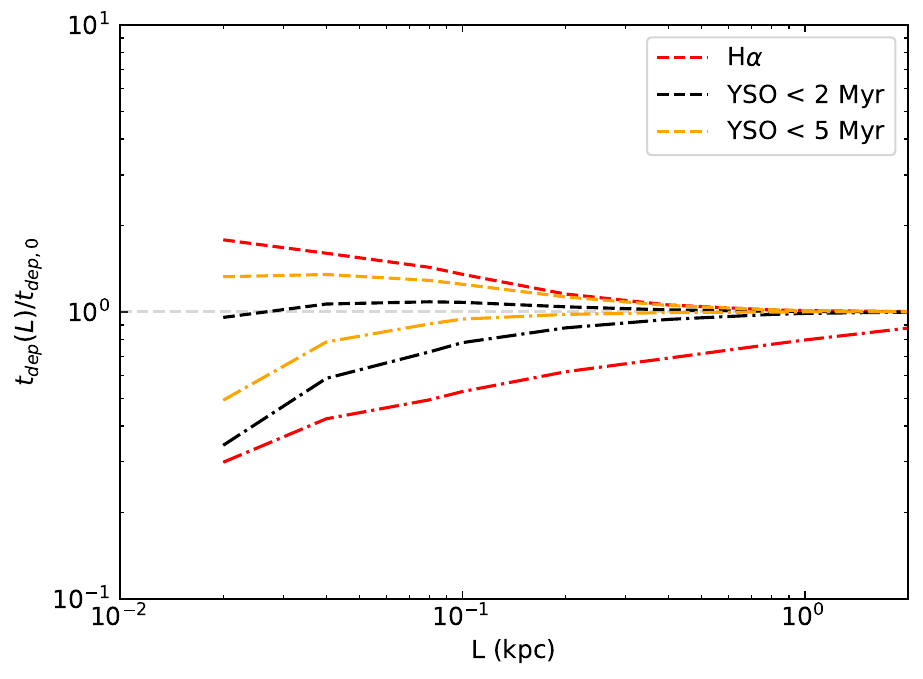}
    \caption{Same as \autoref{fig: tf plot}, but now showing tuning forks derived from SFR maps obtained from YSOs with ages $<2$ Myr (black) and $<5$ Myr (orange) as well as from H$\alpha$ (red). The lines shown are the medians over 20 time snapshots as in \autoref{fig: tf plot}, but we do not show the corresponding 16th to 84th percentile bands for clarity. Different line styles indicate depletion time ratios calculated around CO emission peaks (dashed) or H $\alpha$ emission peaks (dot-dashed).}
    \label{fig: tf plot YSO}
\end{figure}

The fact that H$\alpha$-based SFRs are significantly distorted on small scales by photon transport effects has important implications for the interpretation of the small-scale gas-SFR decorrelation, which is often used to deduce properties of the molecular cloud life cycle and star formation process. To understand these implications, we begin by quantifying how photon transport effects contribute to the tuning fork diagram on small scales. We do this by constructing tuning fork diagrams using the same analysis method described in \autoref{sec: tuning fork diagram}, but this time using the SFR map derived from YSOs instead of H$\alpha$, both for YSOs with ages $<2$ Myr and $<5$ Myr. We show the result in \autoref{fig: tf plot YSO}, together with H$\alpha$-based result we obtained in \autoref{sec: tuning fork diagram}.

We find that, unlike H$\alpha$, YSOs with ages $<2$ Myr are spatially well correlated with the CO emission, so the upper branch of the black plot is close to unity across different scales. The lower branch is below unity, since when the aperture is located around YSO mass peaks, the local stellar particles are grouped together more tightly than the corresponding gas, making the local depletion time smaller than the galactic average value. The comparison between H$\alpha$ and YSOs with ages $<5$ Myr is perhaps more informative, since in this case we are probing very close to the same time scale as H$\alpha$, and thus any differences between the H$\alpha$ and 5 Myr YSO tuning forks translate directly to changes in the inferred properties of molecular clouds. We find that, when the stellar age increases, either YSOs drift away from the CO emission peaks or molecular clouds are dispersed by feedback, resulting in both the upper and lower tuning fork branches deviating from unity. However, the deviation is noticeably smaller than for H$\alpha$.

Examining published analyses based on the SFR-gas decorrelation \citep[e.g.,][]{Kruijssen_2018, Kruijssen_2019, Chevance_2020, Chevance_2021}, it is clear that a tuning fork that opens more narrowly and at small scales implies molecular clouds whose positions are correlated on smaller scales and which spend significantly longer in the ``overlap'' phase when both tracers of star formation and gas are present, and clouds have not been dispersed by feedback; it is likely that the inferred overall molecular cloud lifetime would increase as well.

To quantify this effect, we use the publicly-available \textsc{heisenberg} code \citep{Kruijssen_2018} to analyze all three tuning fork diagrams in \autoref{fig: tf plot YSO}. As a first step in this analysis, we test whether \textsc{heisenberg}'s diffuse emission filter can remove the effects of H$\alpha$ emission coming from diffuse ionised gas (DIG) far from YSOs. The full algorithm implemented by this filter is described by \citet{Hygate_2019}, so here we only summarize for reader convenience. The first step is to first draw a tuning fork diagram and fit the average spacing between emission peaks $\lambda$ from the original emission map; doing so we obtain $\lambda=103.2_{-19.7}^{58.3}$~pc. We then filter out the diffuse component by transforming the H$\alpha$ map into Fourier space, applying a high-pass filter to remove modes with wave number $\gtrsim 1/k$, and then transforming back to physical space. We then re-fit $\lambda$ by applying \textsc{heisenberg} to the filtered map, and iterate the process of fitting, filtering, and re-fitting until $\lambda$ changes by less than 5\% between two successive iterations. Applying this procedure to our H$\alpha$ map, we find that the process converges in six iterations, with the characteristic separation stablizing at $\lambda = 91^{+13}_{-11}$~pc; the mean is only 10\% less than the non-filtered result, and the two are consistent within the formal uncertainties. Moreover, the fraction of diffuse H$\alpha$ flux is measured to be only $\sim 8\%$. These indicate that the effect of DIG is limited in our synthetic observations, a result that can be understood as coming from two effects: first, we have already suppressed much of the DIG emission by removing H$\alpha$ emission below the detection limit; second, the spatially extended H$\alpha$ emission due to photon transmission effect is not distinct from compact \HII regions, as they are originally generated by the same sources. Given the minimal effects of filtering, we turn off this module for the remainder of our analysis with \textsc{heisenberg}, which allows us to use the same procedure to analyze both the H$\alpha$-derived and YSO-derived star formation maps, eliminating a possible source of systematic difference.

We next use \textsc{heisenberg} to derive the durations of two phases in the cloud life cycle: the gas-only phase $t_\mathrm{gas}$ during which only the molecular gas tracer is present and the ``overlap'' phase $t_\mathrm{over}$ during which both tracers of star formation and gas are present. We do this for both the H$\alpha$ map and the two YSO-derived maps. When fitting the durations of these two phases, \textsc{heisenberg} requires that we supply a reference duration of the star-only phase $t_\mathrm{ref}$. For H$\alpha$ emission, we adopt $t_\mathrm{ref} = 4.3^{+0.1}_{-0.2}$ as measured from a disc galaxy simulation at Solar metallicity \citep{Hygate_2019}. For the star formation maps derived from YSOs with ages $<2$~Myr and $<5$~Myr, we set $t_\mathrm{ref}$ to 2 and 5~Myr, respectively. We report the results returned by \textsc{Heisenberg} in \autoref{tab: heisenberg results}.

\begin{table}
    \centering
    \renewcommand{\arraystretch}{1.5} %
    \begin{tabular}{ccccc}
    \hline
    Tracer & $\lambda$ & $t_\mathrm{gas}$ & $t_\mathrm{over}$  & $t_\mathrm{ref}$\\
      & (pc) & (Myr) & (Myr) & (Myr)\\
    \hline
    H$\alpha$ & $103.2^{+58.3}_{-19.7}$ & $27.6^{+15.8}_{-4.7}$ & $8.4^{+6.0}_{-2.0}$ & $4.3^{+0.1}_{-0.2}$\\
    YSO < 5 Myr & $51.1^{+19.3}_{-8.6}$ & $33.2^{+31.6}_{-7.9}$ & $10.6^{+12.4}_{-3.3}$ & $5.0$\\
    YSO < 2 Myr & $38.5^{+10.8}_{-5.9}$ & $141.7^{+33.0}_{-48.4}$ & $30.2^{+8.7}_{-10.5}$ & $2.0$\\
    \hline
    \end{tabular}
\caption{\textsc{heisenberg} fitting results for the tuning fork diagrams shown in \autoref{fig: tf plot YSO}. The first column lists the SFR tracer used, the middle three columns are the fitted physical quantities describing mean cloud separation, duration of gas-only phase, and duration of gas-star overlap phase, and the last column provides the star-only phase duration $t_\mathrm{ref}$, which is used as the reference time scale in fitting.}
    \label{tab: heisenberg results}
\end{table}

From this table it is obvious that when the SFR is traced by YSOs with ages $< 5$~Myr instead of H$\alpha$ emission, the fitted value of $\lambda$ decreases by $\sim 50\%$, and the duration of the ``overlap'' phase increases by $\sim 25\%$. The effect is even more dramatic if we consider YSOs with ages $<2$ Myr, but the uncertainties are extremely large, and it is possible that the method itself breaks down in this case because for these young YSOs the upper branch of the tuning fork does not open at all, contradicting a basic assumption of the method.
Nonetheless, this analysis makes clear that the very short overlap time inferred by previous studies is substantially biased by photon transport effects in H$\alpha$, and that molecular cloud dispersal is perhaps not quite as fast as had been supposed. It is worth noting that this result helps resolve a longstanding tension: existing measurements from CO-H$\alpha$ decorrelation seem to suggest that molecular clouds have an extended quiescent phase when there are no massive stars present, followed by a very rapid dispersal phase once such stars appear, but this is hard to reconcile with the observation that starless molecular clouds in the Milky Way are exceedingly rare. Our finding suggests a resolution: the dispersal time has been significantly underestimated because much of the CO-H$\alpha$ decorrelation is due not to cloud dispersal, but to ionizing photons produced by stars at ages $\approx 2-5$ Myr traveling far from their parent stars before being absorbed and giving rise to H$\alpha$ emission. This picture is also consistent with the recent discovery by \citet{Calzetti23a} using \textit{James Webb Space Telescope} (JWST) of a substantial population of star clusters with ages $\gtrsim 5$ Myr that are still highly dust-embedded.

\subsection{SFR calibration model}
\label{sec: calibration}
In \autoref{sec: why HII fail}, we find that in regions where YSOs are present in regions of high gas column density, H$\alpha$ luminosity does approximately trace the recent star formation rate in the 100-pc scale. However, direct identification of extragalactic YSOs is unavailable for all but the few nearest galaxies due to resolution limits, making it impossible to directly apply such criteria to study cloud star formation rate with H$\alpha$ emission. Nevertheless, calibration of the H$\alpha$ tracer to YSOs is still possible because of the strong correlation between molecular clouds and star formation. In essence, if the cloud mass and area is large enough, we can safely assume the existence of embedded YSOs, and therefore assume that we are looking at a region where H$\alpha$ will be reliable. In our synthetic observations, every CO contour with a mean radius larger than 100 pc has YSOs within it, so we select such CO contours and calculate the total SFR within each of them using both H$\alpha$ and YSOs. Via this process we obtain a sample of 464 distinct CO contours, with a minimum total molecular gas mass of $\sim 10^6 \; \rm M_\odot$ and a minimum SFR$_{\rm H\alpha}$ of $\sim 10^{-3} \; \rm M_\odot yr^{-1}$. The latter limit is also helpful, because this corresponds to an ionising photon flux of $\sim 10^{49}$ yr$^{-1}$, about the luminosity of a single O type star. Therefore, this is roughly also the limit below which we expect stochasticity to foil measurements of the SFR from H$\alpha$ even with limited photon transport effects -- an intuition confirmed by detailed \textsc{slug} simulations by \citet{da-Silva14b}. Thus by selecting contiguous regions of significant CO emission with radii $> 100$ pc, we ensure that these regions almost certainly contain YSOs, and almost certainly contain enough massive stars that stochastic IMF sampling has minimal effects.

\begin{figure}
    \includegraphics[width = \linewidth]{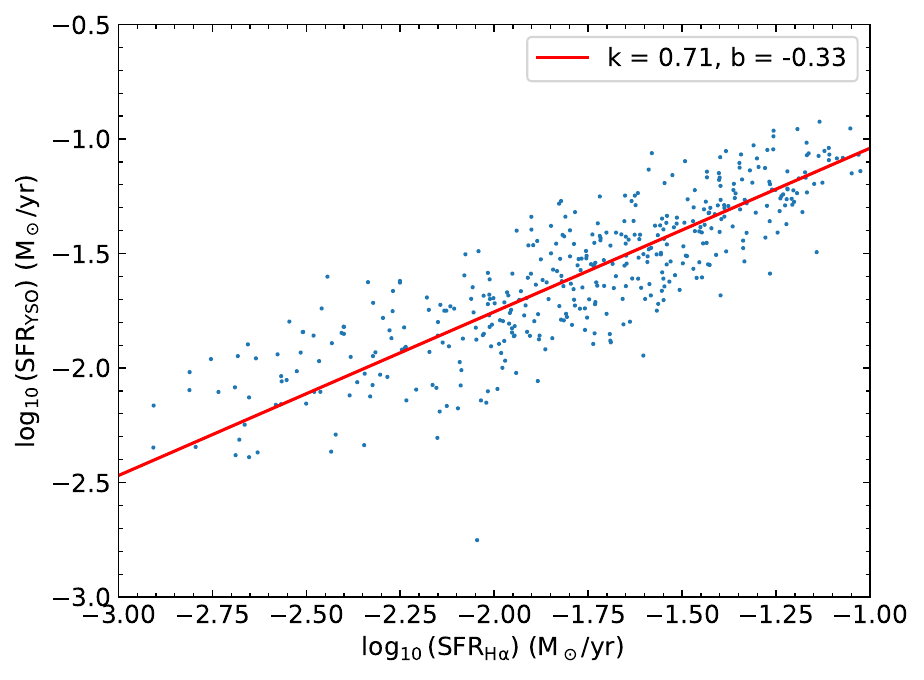}
    \caption{Ratios between SFRs traced by H$\alpha$ luminosity and YSO versus the column density $N_\mathrm{H}$, both calculated from 100-pc-large pixels with YSO inside. The results are stacked from 20 snapshots between 704 Myr and 723 Myr. The mid solid line is the median value, while the shaded region shows the range between the 16th and 84th percentiles.}
    \label{fig: SFR calibration}
\end{figure}

We plot the measured SFR$_\mathrm{YSO}$ vs. SFR$_\mathrm{H\alpha}$ for our contours in \autoref{fig: SFR calibration}. We find that the relation between the two tracers is well-fit by a simple linear (on log scale) model
\begin{equation}
\log_{10} \frac{\mathrm{SFR_{YSO}}}{\mathrm{M}_\odot\;\mathrm{yr}^{-1}} = k \log_{10} \frac{\mathrm{SFR_{H\alpha}}}{\mathrm{M}_\odot\;\mathrm{yr}^{-1}} + b.
\label{eq: SFR calibration}
\end{equation}
The fit coefficients are $k = 0.71$ and $b = -0.33$, and the $R^2$ value is 0.81, indicating a strong linear relation. This relation can be applied in observation to calibrate the star formation rate of individual clouds where a local YSO catalog is unavailable. Although this relation is limited to large molecular clouds, this is not a severe limitation for extragalactic studies, which commonly achieve spatial resolutions of only $\approx 100$ pc \citep[e.g.,][]{Chevance_2020}, close to our cloud radius limit. Note that we did not consider the effect of dust attenuation in our post-analysis, we assume that we observe the total H$\alpha$ emission. For application to real observations one must first dust-correct the H$\alpha$ luminosity \citep[e.g.,][]{Kennicutt_2012,Tacchella_2022}.

\subsection{Future work on nearby galaxies}
\label{sec: future work}
The results in this work are derived from a simulation of a Milky Way-like galaxy. It is unclear to what extent these results will differ for galaxies of very different star formation rates and sizes, where different stellar feedback effects may dominate compared to those in Milky Way-like systems; the properties of the interstellar medium and dust may also differ between galaxies, which can also affect the transmission of ionizing photons.

One approach to mitigate this is of course to extend our simulations to different galaxy types. However, it is also essential to test our results by comparing real high resolution H$\alpha$ luminosity maps with YSO distribution maps. At present the only viable options for such a test are the ``Local Group'' galaxies, including the Large and Small Magellanic Clouds (LMC and SMC), and the M33. H$\alpha$ maps of these galaxies are already available at sub-arcsec resolution, corresponding to physical scale of $\sim 0.2$ pc. The limiting factor is YSOs: the current YSO catalog for the LMC, for example, is still limited to several small regions. Using \textit{Spitzer} IR survey, previous works have \citep{Indebetouw_2008, Chen_2009, Chen_2010} identified YSOs from the LMC's molecular ridge and two \HII complexes. However, the observed area in total is less than 1 kpc$^2$, and thus not sufficient for systematic comparison of H$\alpha$ and YSOs across scales. Moreover, their YSO mass detection lower limit varies between 3 M$_\odot$ and 8 M$_\odot$, making SFR estimates strongly depend on both complex completeness calculations and assumptions about the shape of underlying IMF, especially at the low mass end. Nevertheless, high resolution observations empowered by \textit{JWST} now have great potential to identify extragalactic YSOs. \citet{Peltonen_2024} observe one spiral arm of M33, and find 793 YSOs with a lower mass limit of $\sim 6$ M$_\mathrm{\odot}$. For closer galaxies like the LMC, \textit{JWST} should be able to detect YSOs down to sub-Solar masses. If identified from the whole galaxy, these YSO catalogs will enable a repetition of our simulation analysis here on real data from a suite of nearby galaxies, which can help us better understand the effect of different galactic environments on GMC life cycles. In the future, we plan to run simulations of these galaxies suitable for direct comparisons to such observations.

\section{Conclusion}
\label{sec: conclusion}

In this work, we investigate the discrepancy between estimates of galaxy star formation rates based on H$\alpha$ emission and direct measurements from counts of young stellar objects (YSOs), on spatial scales from a whole galaxy to smaller than individual molecular clouds. To this end we first perform high resolution magneto-hydrodynamic (MHD) simulations of a Milky Way-like galaxy that can resolve cloud-scale structures, with realistic star formation and stellar feedback algorithms, in particular including a treatment of photoionization feedback that allows us to track the locations of ionized gas. Then we post-process the simulation outputs to derive local chemical compositions, non-LTE level populations, and emissivities of both H$\alpha$ and CO $J = 1 \rightarrow 0$ emission lines. We use these to produce synthetic H$\alpha$ and CO emission maps, and compare the former to maps of the true star formation rate as traced by YSOs born with ages $<2$ Myr. We confirm the reliability of our stellar feedback treatment by showing that our simulations reproduce the so-called ``tuning fork'' diagram for observed Milky Way-like galaxies, which quantifies the degree of small-scale decorrelation between CO and H$\alpha$ maps of galaxies.

We find that the H$\alpha$- and YSO-based SFR maps have good agreement on kpc scales, but exhibit significant spatial mismatch on sub-100 pc scales. We show that the primary cause for this discrepancy is the leakage of ionizing photons from \HII regions, especially in low column density regions, with a subdominant contribution due to stars with ages of $\approx 2-5$ Myr drifting away from their birth sites. We therefore conclude that on sub-100 pc scales, H$\alpha$ should be used as a star formation tracer only with extreme caution, and we show that tuning fork diagrams derived from true YSO positions rather than H$\alpha$ emission differ significantly in a way that may have led prior analyses to substantially underestimate the time it takes for newborn stars to disperse their parent clouds.

However, we also show that H$\alpha$ emission does become a reasonably reliable tracer of star formation in regions of high gas density, $N_\mathrm{H} \gtrsim 3 \times 10^{21}$ cm$^{-2}$ when averaged over 100 pc scales, because in such regions ionizing photons are absorbed close to their emission sites and leakage has minimal effects. We propose a calibration model that exploits this result and allows reliable measurements of the SFRs of large molecular regions ($r\gtrsim 100$ pc) from H$\alpha$ luminosity. Future high-resolution observations powered by \textit{JWST} are essential to test our findings and refine this calibration model, and allow more accurate measurements of star formation rates across different environments.

\section*{Acknowledgements}

We thank the anonymous referee for a helpful and constructive
report that improved this paper. We acknowledge Dr. Diederik Kruijssen and Dr. M\'elanie Chevance for suggestions on running the \textsc{heisenberg} code. The authors acknowledge funding from the Australian Research Council through its Future Fellowship and Laureate Fellowship funding schemes, awards FT180100375 and FL220100020, as well as its Discovery Projects scheme (award~DP230102280), the Australia-Germany Joint Research Cooperation Scheme (UA-DAAD), and high-performance computing resources provided by the Australian National Computational Infrastructure (grants~jh2 and ek9) and the Pawsey Supercomputing Centre (grant~pawsey0810) through the National and ANU Computational Merit Allocation Schemes, and by the Leibniz Rechenzentrum and the Gauss Centre for Supercomputing (grants~pr32lo, pr48pi and GCS Large-scale project~10391).

\section*{Data Availability}
 
The data underlying this article will be shared upon reasonable request to the corresponding author.



\bibliographystyle{mnras}
\bibliography{Hu2024} 






\bsp	
\label{lastpage}
\end{document}